\documentclass[aps,prl,superscriptaddress,twocolumn,showpacs]{revtex4-1}
\usepackage{amsmath,amssymb,graphicx,epsfig,color,bbm}
\usepackage[pass]{geometry}
\usepackage[colorlinks=true,citecolor=blue,linkcolor=blue,urlcolor=red,anchorcolor=black,filecolor=black]{hyperref}

\def\nn{\nonumber}

\newcommand{\ba}{\begin{eqnarray}}
\newcommand{\ea}{\end{eqnarray}}

\begin{document}

\title{Photocurrents in Weyl semimetals}

\author{Ching-Kit Chan}
\affiliation{Department of Physics, Massachusetts Institute of Technology, Cambridge, MA 02139, USA}
\author{Netanel H. Lindner}
\affiliation{Physics Department, Technion, 320003 Haifa, Israel}
\author{Gil Refael}
\affiliation{Institute of Quantum Information and Matter and Department of Physics, California Institute of Technology, Pasadena, CA 91125, USA}
\author{Patrick A. Lee}
\affiliation{Department of Physics, Massachusetts Institute of Technology, Cambridge, MA 02139, USA}

\date{\today}

\begin{abstract}
The generation of photocurrent in an ideal two-dimensional Dirac spectrum is symmetry-forbidden. In sharp contrast, we show that three-dimensional Weyl semimetals can generically support significant photocurrents due to the combination of inversion symmetry breaking and finite tilts of the Weyl spectra. Symmetry properties, chirality relations and various dependences of this photovoltaic effect on the system and the light source are explored in details. Our results suggest that noncentrosymmetric Weyl materials can be advantageously applied to room temperature detections of mid- and far-infrared radiations.
\end{abstract}


\maketitle

\begin{figure}[b]
\begin{center}
\includegraphics[angle=0, width=1\columnwidth]{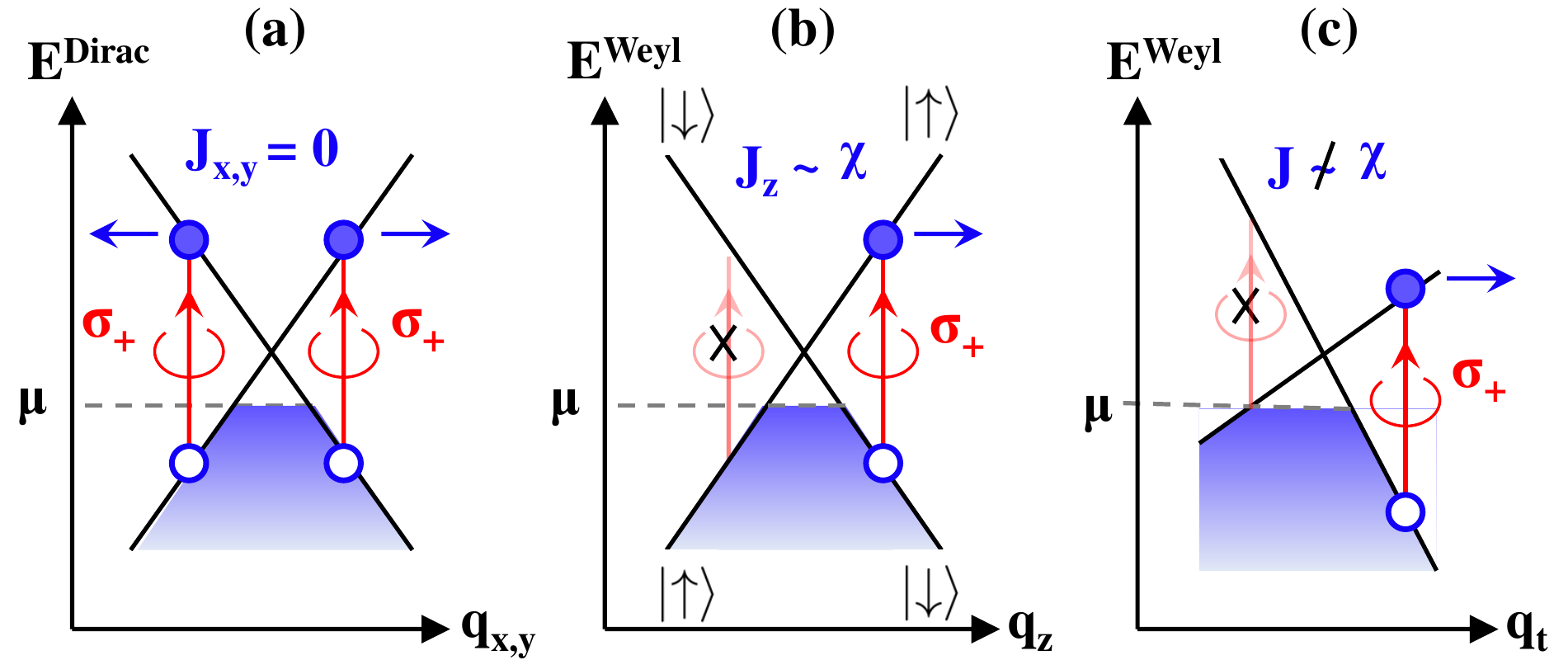}
\caption{Schematics of photocurrent generations in Dirac and Weyl systems. Circularly polarized photons propagating along the z-axis induce spin-flip vertical transitions denoted by the red arrows. (a) In an ideal 2D Dirac system, the excitations are symmetric about the node and thus the photocurrent vanishes. (b) In a 3D Weyl system with an upright crossing spectrum, the extra dimension allows an asymmetric particle-hole excitation along $\rm q_z$ and creates a chirality-dependent photocurrent from each Weyl cone. However, the chiral currents from a monopole and an anti-monopole negate each other, yielding no net current. (c) In the presence of tilt along some direction $\rm q_t$, asymmetric excitations can happen when the system is doped away from the neutrality. The resultant photocurrent is not just determined by the node chirality and the total current is generically non-zero.}
\label{fig_tilt}
\end{center}
\end{figure}

\textit{Introduction.}---Electronic materials with band crossing excitations have recently attracted much interests in condensed matter physics. A 2D Dirac spectrum describes the surface states of 3D topological insulators~\cite{RevModPhys.82.3045,RevModPhys.83.1057} and bulk excitations of graphene~\cite{RevModPhys.81.109}. Their gapless and topological characters have stimulated many electronic applications, one of which is the photovoltaic effect. The linearly crossing dispersions of Dirac systems can absorb photons with, ideally, arbitrarily long wavelength, making them possibly advantageous for infrared (IR) detections. Nevertheless, the generation of photocurrent, defined as the spontaneous production of current without any applied voltage by the exposure to light, has to vanish for an ideal Dirac spectrum in 2D because of the symmetric excitations about the Dirac point [Fig.~\ref{fig_tilt}(a)]. In fact, it has been shown that the resultant photocurrent is negligible even if realistic perturbations including band curvatures, warpings and Zeeman couplings are taken into account~\cite{PhysRevB.83.035309,PhysRevB.88.075144}. So far, the generation of a substantial photocurrent in Dirac systems has to involve external assistances such as couplings to magnetic superlattices~ \cite{2014arXiv1403.0010L}. (Similarly, quantum wires require external magnetic fields to create sizeable photocurrents~\cite{PhysRevB.85.085311}). Producing photocurrents in Dirac materials remains challenging and has been an active research subject~ \cite{W.2012,PhysRevB.92.241406,PhysRevB.93.125125,2016arXiv160703888K,PhysRevLett.116.076801,PhysRevB.93.081403,2014arXiv1403.0010L}.

In this paper, we propose that Weyl semimetals can generically develop photocurrents without the need of external couplings. Weyl spectrum is the 3D generalizations of the Dirac cone and thereby shares the same advantage of long-wavelength photon absorptions. Unlike Dirac systems, Weyl semimetals necessarily break either time-reversal (TR) symmetry or spatial inversion (I) symmetry, or both. The photocurrent response of a Weyl system differs from the Dirac counterpart in two crucial ways. First, Weyl cones have definite chiralities and always come in a pairs. They can be regarded as topological monopoles or anti-monopoles of the Berry curvature. For an upright Weyl cone [Fig.~\ref{fig_tilt}(b)], the absorption of a circularly polarized photon flips the spin, resulting in asymmetric excitations along the drive direction. Yet, the direction of the photocurrent is governed by the chirality and hence, the sum of photocurrents from a Weyl node pair has to vanish identically. On the other hand, a Weyl cone can be tilted~\cite{PhysRevB.91.115135,Soluyanov2015} because of reduced symmetries. The corresponding photoexcitation is highly asymmetric about the nodal point [Fig.~\ref{fig_tilt}(c)]. The consequential photocurrent is controlled by the tilt and the chirality and there is generally no offset between photocurrents unless additional symmetries are imposed. Note that other interesting effects such as gyrotropic magnetic effect~\cite{PhysRevLett.116.077201,PhysRevLett.115.117403}, photovoltaic chiral magnetic effect~ \cite{PhysRevB.93.201202}, anomalous Hall effect~\cite{chan16}, emergent electromagnetic induction~\cite{2016arXiv160706537I} and nonlinear optical responses~ \cite{2016arXiv160904894W} can occur in Weyl semimetals, but have different physical origins.

Photocurrents in systems without I symmetry have been observed in semiconductor quantum wells~\cite{Ganicheva2001} and tellurium~\cite{ivchenkoandpikus}. A variety of mechanism has been discussed~\cite{ivchenkoandpikus}, including real and virtual absorptions and spin-dependent scatterings~\cite{Ganicheva2001}, but the effect is small and the discussion has been limited to quadratic band structures. Here we show that the linear dispersion in Weyl semimetals hold a special advantage and a large photocurrent proportional to the absorption can be produced which survives up to room temperature. In addition to the broken I-symmetry, we find that the presence of finite tilts of the Weyl dispersions, being commonplace in realistic materials, is the key for the photocurrent in Weyl semimetals. The importance of the tilt has not been discussed in the literature so far. The photocurrent response does not require the chemical potential to be tuned to the Weyl point, nor any imbalance of chemical potentials between opposite Weyl nodes, which are otherwise important for chiral-anomaly related responses~\cite{vafek14,0953-8984-27-11-113201,2016arXiv160505409M,PhysRevB.93.201202}. We investigate the conditions and magnitudes of photocurrents induced by a laser drive and further discuss the potential application for room temperature detections of blackbody IR radiations.

\textit{Photocurrent response.}---We start by discussing the photocurrents generated from a single Weyl spectrum without any symmetry restriction. The low-energy effective Hamiltonian can be generally written as:
\begin{eqnarray}
\label{eq_H}
H_W(\vec q)=\hbar v_t q_t \sigma_0 + \hbar v_F \hat v_{i,j}q_i \sigma_j,
\end{eqnarray}
where $\rm v_F$ is the Fermi velocity without tilt and $\rm \sigma_j$ are Pauli matrices. $\rm \hat v_{i,j}$ represents anisotropy and $\rm \chi= Det \left(\hat v_{i,j}\right) = \pm 1$ determines the chirality.   $\rm v_t$ gives the tilt velocity and $q_t = \hat t \cdot \vec q$ with $\hat t$ being the tilt direction. The linear Weyl dispersion is given by $E_\pm (\vec q) = T(\vec q)\pm U(\vec q) =\hbar v_t q_t \pm \hbar v_F [ \sum_j (\sum_i \hat v_{i,j}q_i)^2 ]^{1/2}$ and the ratio $\rm v_t/v_F$ measures the tilt. When $|T(\vec q)/U(\vec q)|$ is less than $\rm 1$ for all $\rm \hat q$, the node is in the type-I phase, whereas when it is greater than $\rm 1$ for some $\hat q$, the system is in the type-II regime, in which electron and hole pockets are formed~\cite{Soluyanov2015}.

The interaction with a monochromatic light characterized by $\vec A(t)=\vec A_{+} e^{- i \omega t} + \vec A_{-} e^{ i \omega t}$ enters through the Peierls substitution, leading to the interaction Hamiltonian $\rm V(t)=V_+ e^{- i \omega t} + V_- e^{ i \omega t}$ with
\begin{eqnarray}
V_\pm = \hbar v_F \hat v_{i,j} A_{\pm,i} \sigma_j.
\end{eqnarray}
$\rm V_{+(-)}$ describes the spin-dependent photon absorption (emission) process. In the isotropic limit $\rm \hat v_{i,j}=\delta_{i,j}$, a circularly polarized light propagating along $\rm q_z$ corresponds to $\rm V_\pm=\rm \hbar v_F A \sigma_\pm/2$. We have ignored Zeeman coupling here because the ratio of Zeeman to orbital couplings is $\rm \sim g_s \hbar  \omega/(2 m v_F c) \sim  10^{-3}$~\cite{PhysRevB.88.075144} based on reported g-factors~\cite{Hu2016,Jeon2014}.

\begin{figure}[t]
\begin{center}
\includegraphics[angle=0, width=1\columnwidth]{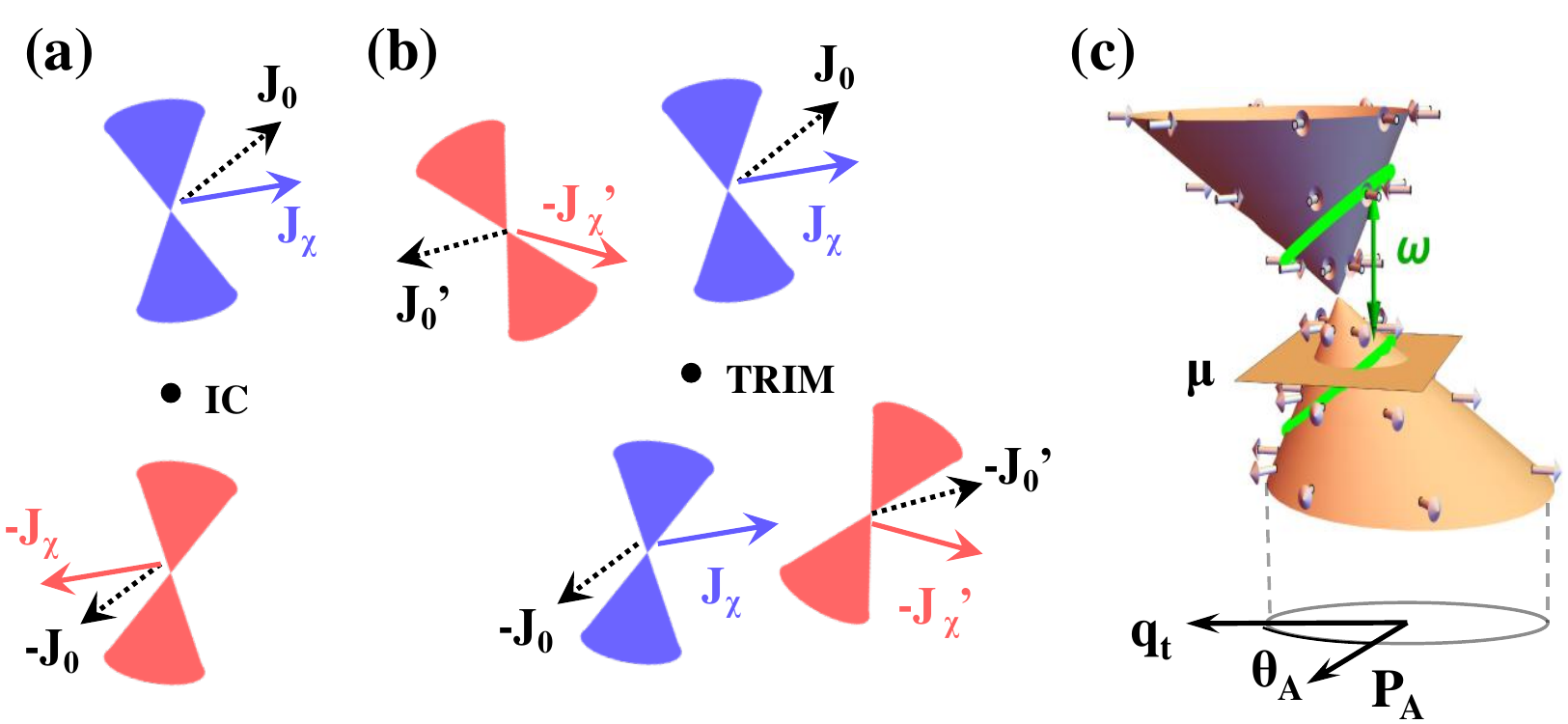}
\caption{Relations between photocurrents in centrosymmetric and noncentrosymmetric Weyl semimetals. Each Weyl node produces a photocurrent with chiral-independent ($\vec J_0$) and chirality-dependent ($\vec J_\chi$) components. (a) In the presence of an inversion center, a pair of I-related Weyl nodes have opposite tilt and opposite chirality, leading to the cancelation of photocurrents. (b) In noncentrosymmetric Weyl systems, TR symmetry relates two Weyl nodes of the same chirality instead. Monopoles and anti-monopoles are not symmetry-related and can have different tilts. Two pairs of Weyl nodes give rise to an overall photocurrent of $2( \vec J_\chi - \vec J_\chi')$. (c) defines the angle $\rm \theta_A$ between the tilt $\rm q_t$ and Poynting vector $\rm P_A$.}
\label{fig_inversion}
\end{center}
\end{figure}

We now compute the photovoltaic current $\vec J = (-e) \sum_{q,l=\pm} \left[\partial E_l(\vec q)/\partial \vec q\right] \times \left[n_l(\vec q) -n_l^0 (\vec q)\right]$, where $n_\pm(\vec q)$ and $n^0_\pm(\vec q)$ are the perturbed and equilibrium distribution functions, respectively. Within Fermi's golden rule and the relaxation time approximation, each Weyl node contributes a photocurrent density:
\begin{eqnarray}
\label{eq_J}
J_i= \left(\frac{-e \tau \omega^2 A^2}{16 \pi^2}\right) \bar J_i,
\end{eqnarray}
with a dimensionless response function
\begin{eqnarray}
\label{eq_Jbar}
\bar J_i(\omega)&=& 4 \int d^3\left(\frac{v_F q}{\omega}\right) \frac{\partial [\Delta E(\vec q)/\hbar]}{\partial (v_F q_i)} \left| \left\langle q_+ \left| \frac{V_+}{\hbar v_F A}\right|q_-\right\rangle \right|^2 \nn \\
&&\ \ \ \ \ \  \times \delta\left( \frac{\Delta E(\vec q)}{\hbar \omega}-1\right) \left[n_-^0(\vec q) - n_+^0(\vec q) \right],
\end{eqnarray}
where $\rm \Delta E = E_+ -E_-$. We have introduced a relaxation time $\rm \tau$ to account for disorder and phonon scattering~ \cite{PhysRevB.84.235126,PhysRevLett.108.046602}. This response function describes the vertical transition from state $\rm \left| q_- \right\rangle$ to $\rm \left| q_+ \right\rangle$ by absorbing a photon with frequency $\omega$. Each particle-hole excitation produces a current $ -e \partial [\Delta E(\vec q)/\hbar]/\partial \vec q$ being independent of the tilt. We note that Eq.~(\ref{eq_Jbar}) is a dimensionless number which depends on the tilt $\rm v_t/v_F$ but is independent of $\rm v_F$ for a given tilt. Recently it was shown that for a single node with negligible tilt, the trace of the response (i.e. $\sum_{P_A=\{x,y,z\}}\bar J_{i=P_A}$ where $\rm P_A$ is the Poyting vector) is universal and proportional to the chirality~\cite{2016arXiv161105887D}. It is easy to see that this result survives for finite tilt over a limited range of chemical potential.


For Weyl semimetals, it is possible to break both TR and I symmetries. In these cases, each Weyl node can have different parameter values and chemical potentials. $\rm \bar J$ coming from different nodes are not symmetry-related and there is no current cancelation. Only when there is I symmetry, as we show below, will the photocurrents lead to cancelation.

\textit{Symmetry consideration.}---In centrosymmetric Weyl semimetals, a Weyl node at $\vec k$ is related to another one at $-\vec k$ about the inversion center (IC) [Fig.~\ref{fig_inversion}(a)]. Their Hamiltonians take the same form as Eq.~(\ref{eq_H}) with the relations $\rm q_i \leftrightarrow -q_i$ and $\rm \sigma_j \leftrightarrow \sigma_j' = P \sigma_j P^{-1}$. The inversion $\rm P$ just changes the basis of $\sigma_j$. Hence, I-related nodes have opposite tilt and opposite chirality. Similarly, with TR symmetry, two Weyl nodes are related about the time-reversal invariant momentum (TRIM) with the relations $\rm q_i \leftrightarrow -q_i$ and $\rm \sigma_j \leftrightarrow -\sigma_j'' = T \sigma_j T^{-1}$, where $\rm T$ is the TR transformation. Thus, two TR-related nodes have opposite tilt but the same chirality [Fig.~\ref{fig_inversion}(b)]. Even though each monopole has to be accompanied by an anti-monopole, there is no symmetry restriction between them.

Importantly, the photocurrent depends on the sign of the tilt and the chirality. According to the response function [Eq.~(\ref{eq_Jbar})], when we change $\rm v_t \rightarrow -v_t$ and $\rm \hat v_{i,j} \rightarrow -\hat v_{i,j}$, it is equivalent to $\vec q \rightarrow -\vec q$ and the integral yields $\bar J_i \rightarrow - \bar J_i$. With this relation, we can decompose the photocurrent from the n-th Weyl node as
\begin{eqnarray}
\label{eq_Jsymmetry}
\vec J^{(n)} (\text{sgn}[v_t^{(n)}],\chi^{(n)})= \text{sgn}[v_t^{(n)}] \vec J^{(n)}_0 + \chi^{(n)} \vec J^{(n)}_\chi.
\end{eqnarray}
This is because the other two components (one is proportional to $\rm \text{sgn}[v_t^{(n)}] \times \chi^{(n)}$ and the other is independent on any of them) have to sum to zero. In. Eq.~(\ref{eq_Jsymmetry}), the first component changes sign for opposite tilts and can be generated by a linearly polarized drive, whereas the second chirality-dependent term can only be induced by a circularly polarized light.

Mirror symmetry is common in realistic Weyl materials. A mirror reflection flips the sign of the Weyl momentum along the mirror axis and relates a monopole to an anti-monopole. However, it does not lead to cancelation of photocurrent because the drive breaks the mirror symmetry of the photocurrent response.

With these symmetry considerations, we conclude that the overall photocurrent has to balance out when there is I symmetry. For noncentrosymmetric systems with TR symmetry, a minimal number of two pairs of Weyl nodes can produce a non-zero response $2(\vec J_\chi-\vec J_\chi' )$. The photocurrent relations are graphically summarized in Fig.~\ref{fig_inversion}(a-b). In a Weyl system without any symmetry, both $ \vec J_0$ and $ \vec J_\chi$ can survive. Note that the non-zero photocurrent originates from the different tilts and anisotropies between monopoles and anti-monopoles. While any chemical potential difference can help in a similar way, we shall only consider identical chemical potentials for each Weyl node for the rest of the paper.

\textit{Photocurrent by monochromatic drives.}---We explore various dependences of photocurrents in noncentrosymmetric Weyl semimetals with TR symmetry due to a circularly polarized drive. For illustrative purpose, we consider a minimal setup of four Weyl nodes, where two monopoles are tilted along $\rm q_z$ with opposite tilt velocities $\rm v_t$ and $\rm -v_t$, while the anti-monopole dispersions are upright. The photon propagates along the $\rm q_x-q_z$ plane and is parameterized as $\vec A(t) = A \left(\cos\theta_A \cos\omega t, \xi \sin\omega t, -\sin\theta_A \sin \omega t\right)$ with $\xi=\pm 1$ controlling the polarization. Below, we consider the isotropic limit, i.e. $\hat v_{i,j}=\delta_{i,j}$.

\begin{figure}[t]
\begin{center}
\includegraphics[angle=0, width=1\columnwidth]{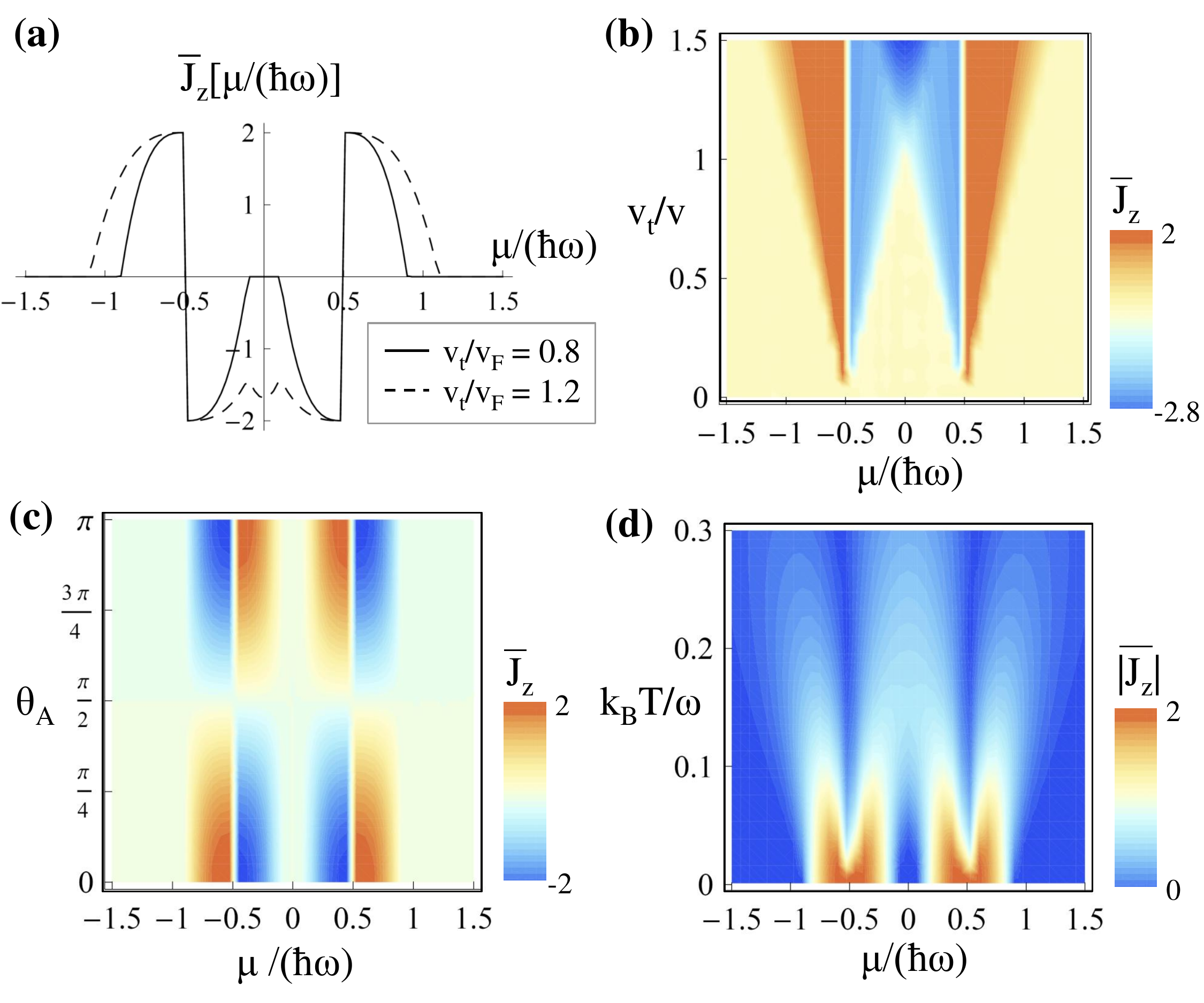}
\caption{Demonstrations of photocurrents in a noncentrosymmetric Weyl semimetal driven by a monochromatic circularly polarized light of frequency $\rm \omega$. Two pairs of isotropic Weyl nodes are considered with tilt velocities $\rm v_{t}$, $\rm -v_{t}$, $0$ and $0$. The tilts are along $\hat z$ and the light propagates at an angle $\rm \theta_A$ from $\rm \hat z$. $\rm \bar J$ gives the dimensionless current. (a) Photocurrent at $\rm T=0$ and $\rm \theta_A=0.3$. With a type-I tilt (i.e. $\rm |v_t| < v_F$), $\rm \bar J_z$ is non-zero when the chemical potential $\rm \mu$ falls in the region that supports asymmetric transitions. The sharp change of $\rm \bar J_z$ at $\rm |\mu|/(\hbar \omega) = 0.5$ is caused by the fact that the two nodes without tilt can no longer be excited when $\rm |\mu| > \hbar \omega/2$. When the tilt is of type-II ($\rm |v_t| > v_F$), $\rm \bar J_z$ does not vanish even at $\mu =0$. (b) $\rm \bar J_z$ as a function of the tilt $\rm v_t/v_F$ and $\rm \mu$ at $\rm T=0$ and $\rm \theta_A=0.3$. The region of non-zero $\rm \bar J_z$ increases with the tilt of the Weyl spectra. (c) $\rm \bar J_z$ as we vary $\theta_A$ at $\rm T=0$ and $\rm v_t/v_F=0.8$. The flow of the photocurrent follows the light propagation. (d) Temperature dependence of $\rm  |\bar J_z|$ at $\rm v_t/v_F=0.8$ and $\rm \theta_A=0.3$. For $\rm \hbar\omega \sim 120~meV$ and at room temperature ($\rm k_B T /(\hbar \omega) \sim 0.22$), $1/4$ of the value still remains.}
\label{fig_photocurrent}
\end{center}
\end{figure}

Figure~\ref{fig_photocurrent}(a) plots the dimensionless photocurrent response $\rm \bar J_z$ at zero temperature with $\theta_A=0.3$ and finite tilts. When $\rm 0 < v_t/v_F <1$, the signal is non-zero due to asymmetric excitations assisted by the tilt [Fig.~\ref{fig_tilt}(c)]. When $\rm  \left |\mu\right| < \left|-v_t/v_F+1 \right| \hbar \omega/2$, the Weyl nodes are symmetrically excited and the chiral photocurrents from the monopoles and the anti-monopoles carry the same magnitudes but opposite directions. As $\rm |\mu|$ is increased, the magnitudes of both currents decrease in a tilt-dependent manner, resulting in an imbalanced photocurrent $ 2 [ \vec J_\chi(\mu/\omega) -\vec J_\chi'(\mu/\omega)]$. When $\rm \left |\mu \right|$ exceeds $\rm \left|v_t/v_F+1 \right| \hbar \omega/2$, vertical transitions are no longer allowed and every current vanishes. Note that the sharp jump of $\bar J_z$ at $\rm|\mu| =\hbar \omega /2$ happens because we consider the special case of two untilted Weyl cones and is not a generic feature.

The region for non-vanishing current can be expanded by increasing the tilt. In the type-II Weyl phase with $\rm |v_t|>v_F$, we have $\bar J_z \neq 0$ as long as $\rm |\mu| < \left|v_t/v_F+1 \right| \hbar \omega/2$. Interestingly, the photocurrent survives at $\rm \mu =0$. The reason is that in the type-II regime, the photoexcitation depends strongly on the direction $\rm \hat q$ and thus is anisotropic even at the neutrality point. Figure~\ref{fig_photocurrent}(b) presents the tilt and chemical potential dependences for the photocurrent. It provides an idea of the working requirements for the photovoltaic effect. $\bar J$ is of order one and its sign is controlled by the light polarization. In the isotropic limit, the photocurrent and the Poynting vector of the drive share the same direction as shown in Fig.~\ref{fig_photocurrent}(c).


In order to investigate the possibility for room temperature photocurrent generation, we plot the temperature dependence of $|\bar J_z|$ in Fig.~\ref{fig_photocurrent}(d). Our results indicate that a sizeable portion of the photocurrent survives with $\rm k_B T \simeq 0.2~\hbar \omega$. For an IR radiation with $\rm \hbar \omega \sim 120~meV$ and room temperature $\rm k_B T = 26~meV$, $\bar J$ remains a quarter of its zero-temperature value.

We provide an estimation of the photocurrent density driven by a CW $\rm CO_2$ laser. For IR detection applications, we consider $\rm \hbar \omega \sim 120~meV$, and a typical laser intensity $\rm I = \epsilon_0 c (\hbar \omega A /e)^2 \sim 10^{6}~W m^{-2}$~\cite{W.2012}. At low temperature, the measured $\rm \tau \sim 45~ps$, corresponding to a long mean free path ($\rm \sim 5~\mu m$)~\cite{2015arXiv150200251Z}. At room temperature, $\rm \tau$ can be reduced by 50 as inferred from resistivity measurements~\cite{2015arXiv150200251Z}. Such a large $\rm \tau$ implies a weak disorder effect in Weyl semimetals. We estimate a large $\rm |J| \sim 4.3 \times 10^{7}~Am^{-2}$ at low temperature. We remark that this result does not require any $\rm \mu$ imbalance. Instead, finite tilts and $\mu$ are necessary.

Material candidates to probe this photovoltaic effect include the type-I TaAs family~\cite{Huang2015,PhysRevX.5.011029,Xu613,PhysRevX.5.031013}, or the type-II compounds such as $\rm WTe_2$~\cite{Soluyanov2015} and $\rm MoTe_2$~\cite{2015arXiv151107440W}. For type-I materials with small tilts, one would require $\rm \hbar \omega \sim 2\mu$. With a finite tilt, for instance $\rm SrSi_2$~\cite{Huang02022016} that has $\rm v_t/v_F\sim 0.6$ and $\rm \mu \sim 22~meV$, one can have a detection window of wavelength $\rm \sim 10-50~\mu m$. The operational window becomes even larger for type-II Weyl semimetals according to Fig.~\ref{fig_photocurrent}(b).

A film geometry is required to avoid any absorption issue. For a system of thickness $\rm L_z$ and area $\rm L_x L_y$ with four Weyl nodes, the number of photons being absorbed ($4 L_xL_yL_z \sum_q [n_+(\vec q)-n_+^0(\vec q)]$) and the number of photons from the drive ($\rm I(\omega) L_xL_y \tau /(\hbar \omega)$) provide an estimation of the absorption length scale $\rm L_a \sim 4 \pi^2 \hbar \epsilon_0 c v_F /(\omega e^2)\sim 700~nm$. An ideal device shall have $\rm L_z \lesssim  L_a$.

We contrast the performance of our effect with other photodetectors using the external quantum efficiency $\rm \eta$, defined as the ratio of the number of charge carriers to the number of incident photons~\cite{0034-4885-77-8-082401}. For a Weyl semimetal with $\rm L_z = 100~nm$ and a lateral dimension of a few $\rm \mu m$, $\eta = \hbar \omega |J| L_z/(e I L_x) \sim 10^{-3}-10^{-2}$ for mid-infrared frequency ($\rm \sim 0.1~eV$) at room temperature. In comparison, graphene grown on substrates designed to break I symmetry has been under study as photodetector. The corresponding substrates introduce a strong disorder scattering and the existing and optimized photodetector has a very low $\rm \eta \sim 10^{-5}$~\cite{6799997}. We also emphasize that in comparison with devices based on photoconductivity, in our case the carriers move with a net velocity, giving rise to a current without any background.

\textit{Blackbody radiation detections.}---Conventional semiconductor-based photodetectors are limited to wavelength less than $\rm 4~\mu m$ due to finite bandgaps~\cite{0034-4885-77-8-082401}. Even in graphene-based devices, the photodetections are either restricted to the near-infrared regime~\cite{Koppens2014}, or suffer from low efficiency~\cite{6799997}. Our proposed effect can overcome these challenges through its wide range of working window. According to Planck's law, a blackbody object at an equilibrium temperature $\rm T_b$ has a continuous spectrum peaked at $\rm \hbar \omega \sim 2.8~k_B T_b$, corresonding to $\rm 73~meV$ (or $\rm  17~\mu m$) at room temperature. By a straightforward generalization of Eq.~(\ref{eq_J}) and~(\ref{eq_Jbar}) for blackbody radiation, one can show that up to $\rm \sim 1/3$ of the dimensionless response survives at room temperature~\cite{note1}, indicating the suitability of Weyl semimetals for IR detections. We remark that broadband quarter-wave plates as circular polarizers exist for both mid-\cite{Sieber:14} and far-infrared~\cite{Masson:06,Chen20131} regimes.

\textit{Conclusion.}---Based on symmetry analyses and perturbative calculations, we predict a significant photocurrent generation in noncentrosymmetric Weyl semimetals. The photovoltaic process is a consequence of broken I symmetry and finite tilts of the dispersions which are unique to Weyl systems. We show that the effect remains significant in a large window of operating parameters, correspondence to temperature, chemical potential, and frequency of the light source. The predicted photocurrent can be readily detected using standard laser experiment techniques and existing Weyl materials. Our findings reveal the suitability for using Weyl semimetals as room temperature IR detectors.

\textit{Acknowledgements.}--- We thank Nuh Gedik for very helpful discussions. P.A.L. acknowledges the support from DOE Grant No. DE-FG02-03-ER46076 and the Simons Fellows Program. G.R. acknowledges the supported from NSF through DMR-1410435. N.H.L. acknowledges supports from I-Core, the Israeli excellence center Circle of Light, the People Programme (Marie Curie Actions) of the European Union's Seventh Framework under REA Grant Agreement No. 631696 and financial support from the European Research Council (ERC) under the European Union's Horizon 2020 research and innovation programme (Grant agreement No. 639172). P.A.L. and C.-K.C. thank the hospitality of the CMT group at Caltech where this work was carried out. G.R. thanks the Aspen Center for Physics where a part of the work was done.

\bibliography{Weyl-photocurrent-refs}

\clearpage
\begin{widetext}

\begin{center}
\large{\bf Supplemental Material:\\ Photocurrents in Weyl semimetals}\\
\vspace{14pt}
\normalsize{Ching-Kit Chan, Netanel H. Lindner, Gil Refael, and Patrick A. Lee}
\end{center}

\section{Room temperature detection of blackbody infrared radiation}

We demonstrate the use of noncentrosymmetric Weyl semimetals as room temperature IR detectors for radiations coming from a blackbody object at an equilibrium temperature $\rm T_b$. We consider the same setup used in Fig.~3 in the main text, that is two pairs of Weyl nodes in the presence of TR symmetry.

A blackbody object held at an equilibrium temperature $\rm T_B$ has a continuous radiation intensity spectrum $ I(\omega,T_B) = \hbar \omega^3 (4\pi^2 c^2 )^{-1} (e^{\hbar \omega/k_B T} - 1)^{-1}$, which is peaked at $\rm \hbar \omega \sim 2.8~k_B T_b$. Generalizing Eq.~(3) and~(4) in the main text to take into account the continuous blackbody spectrum, each Weyl node brings about a photocurrent density:
\begin{eqnarray}
\label{eq_Jblackbody}
J^b_i(T,\mu) &=& \left(\frac{-e^3 \tau k_B^4 T_b^4}{128 \pi^4 \hbar^5 \epsilon_0 c^3}\right) \bar J^b_i \nn \\
 &=&\left(\frac{-e^3 \tau k_B^4 T_b^4}{128 \pi^4 \hbar^5 \epsilon_0 c^3}\right) \int_0^\infty dx \frac{x^3 \bar J_i(\omega=\frac{x k_B T_b}{\hbar})}{e^{x}-1}.
\end{eqnarray}
The prefactor gives the strength of the photocurrent density, while the dimensionless integral, defined by $\rm \bar J^b_i$ above, corresponds to the dimensionless photocurrent response due to a blackbody radiation.

The IR detection performance is presented in Fig.~\ref{fig_blackbody_supple} as a function of temperature for various chemical potential values. At $\rm \mu=0$, the photocurrent vanishes at $\rm T=0$ and steadily increases with $\rm T$ until $\rm T \approx 0.5~ T_b$, in consistent with Fig.~3(d) in the main text. For finite $\rm \mu \approx k_B T_b$, the photocurrent drops monotonically with temperature with up to $\rm 35\%$ of $\rm \bar J^b$ survives at $\rm T=T_b$. This reduction value is comparable to that due to the monochromatic drive shown in Fig.~3(d) of the main text. Note that even for room temperature, the conditions $\rm k_B T \approx k_B T_b \approx \mu \approx 26~meV$ can be easily achieved in realistic Weyl semimetal materials. Our estimation gives a magnitude of $\rm J^b \sim 100 ~Am^{-2}$ at room temperature. All these results illustrate that Weyl semimetals are excellent candidates for IR detectors.

\begin{figure}[h]
\begin{center}
\includegraphics[angle=0, width=0.6\columnwidth]{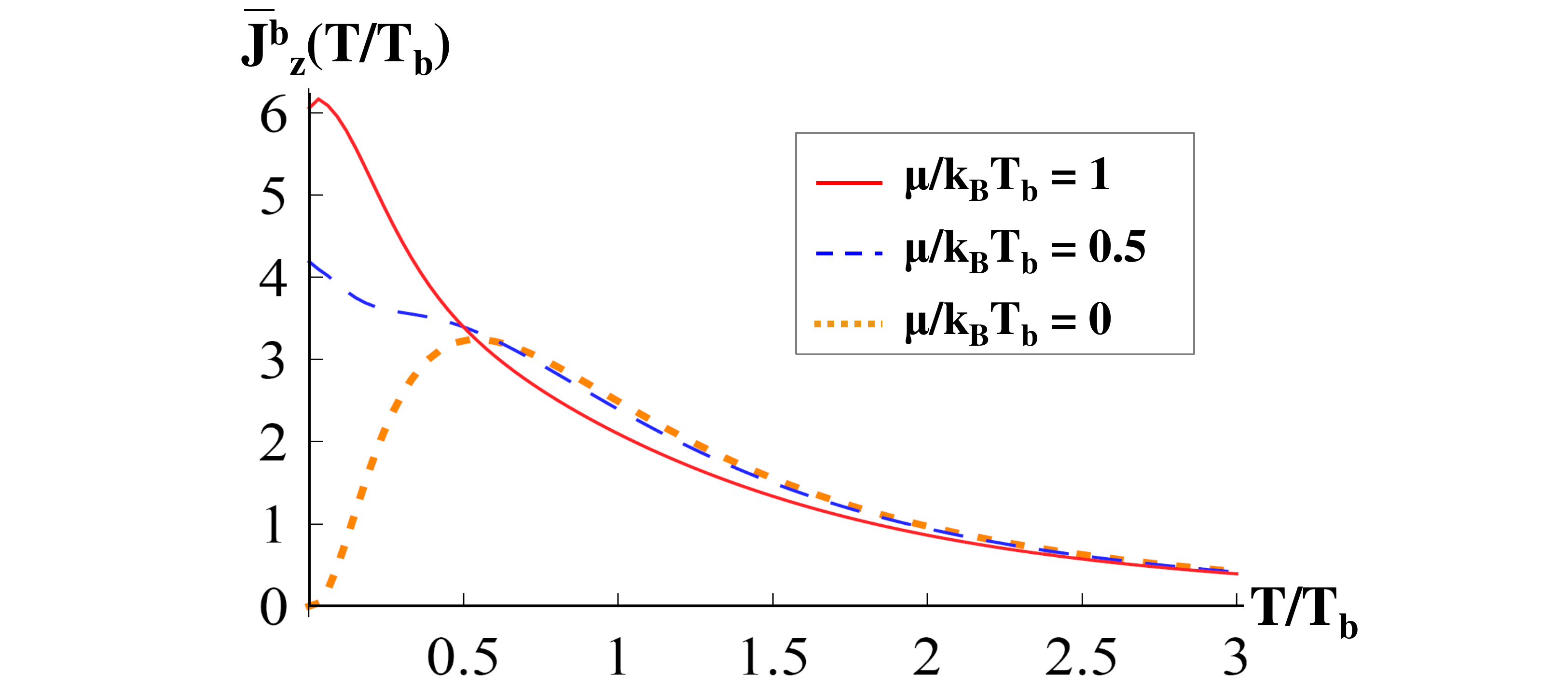}
\caption{Dimensionless photocurrent response $\rm \bar J_z^b$ due to a blackbody radiation with room temperature $\rm T_b$. We take $\rm v_t/v_F=0.8$ and $\rm \theta_A=0$. At $\rm \mu=0$, the photocurrent vanishes at $\rm T=0$ because of symmetric excitations and requires finite temperature to have a non-zero response. When $\mu \sim k_B T_b$, $\rm \bar J_z^b$ just decays with temperature, with $\rm 35\%$ remaining at $\rm T=T_b$.
}
\label{fig_blackbody_supple}
\end{center}
\end{figure}

\end{widetext}

\end{document}